# Determining Hydration Level in Self-Assembled Structures Using Contrast Variation Small Angle Neutron Scattering

*Albert Y. Ho, Guan-Rong Huang, and Wei-Ren Chen*

Neutron Scattering Division, Oak Ridge National Laboratory, Oak Ridge, 37831, Tennessee, United States

**ABSTRACT**

We outline a strategy for quantitatively evaluating the conformational characteristics of self-assembled structures using the techniques of contrast variation small angle neutron scattering. By means of basis function expansion, a case study of spherical micelles demonstrates that the intra-particle hydration and polymer distributions can be determined from the coherent scattering intensity in a model-free manner. Our proposed approach is simple, analytical and does not require a presumptive hypothesis of scattering function as an input in data analysis. The successful implementation of the proposed approach opens the prospect of quantifying the nanoscale complexity of soft matter using neutron scattering.

The most important thermodynamic factor driving small molecules to organize into complex self-assembled structures in aqueous environments is the effect of hydrophobicity.[1] In nature, biopolymers characterized by conformational complexity and functionality form hierarchical structures, such as proteins, ribosomes, cell membranes and organelles.[2] Self-organizing synthetic amphiphiles also offer a thermodynamic pathway for nanomanufacturing to create novel functional materials with a wide variety of sophisticated structures at nanoscales.[3]

The dominant role of water in this effect that underlies the self-assembly process has been long recognized:[1,2] the morphology of self-assembled structures is determined by the free energy change originating from the competition of entropy, which prefers the solubility of molecular building blocks as monomers in water, and the overall bond energy of hydrogen network, which promotes their self-organization.[4] To elucidate the links between the amphiphilic balance inherent to molecular design and micellization phenomenon, fundamental insights into the local hydration levels in self-assembled structures is of vital importance for both theoretical studies and practical purposes.

Prominent among the experimental techniques for accessing the hydration level of micelles is small angle neutron scattering (SANS).[5-6] In this technique, the measured scattering cross section is a collective reflection of the scattering length density (SLD) fluctuations of micellar solutions. While it is clear that the surfactant molecule accounts for a significant portion of coherent intensity due to the compositional difference between the polymer and water, a non-negligible contribution of collected signal is from the density fluctuation caused by the restructuring of invasive water molecules inside micelles.[7-9] Although the chemical composition of the polymer remains unchanged, inside micelles, the spatial arrangement of invasive water molecules around surfactant molecules is different from that of bulk water.[10] This difference in packing also results in a change in SLD. One unique advantage of SANS is its sensitivity to isotopic substitution:[11] By varying the protium–deuterium ratio in water, the spatial correlation from invasive water, critical information about the thermodynamic, and kinetic stability of self-assembled structures can be experimentally compartmentalized without fundamentally altering the intrinsic characteristics of the micellar solution.

To determine the intra-micellar hydration level using the technique of contrast variation SANS, an analytical model of scattering function needs to be *a priori* selected for data analysis. Depending on the individual amphiphilic balance, different surfactants aggregate and hydrate to different extents during micellization, thereby affecting their self-assembled structures at equilibrium. Evidenced by documented studies, there exists a great amount of phenomenological models to account for different scattering behaviors of self-assembly systems with different morphologies.[7-9, 12-23] While these parametric models can conveniently facilitate the analysis of scattering data to provide a quantitative description of micellar structure, they can also often lead to significantly biased conclusions if the wrong conceptual picture of micellar conformation is used as an input in model development.

To circumvent this intrinsic limitation of analytical modeling, in this report we propose a model-free spectral analysis approach, using the tactic of basis function expansion, to determine the micellar hydration and other conformational characteristics. The main conceptual advantage of basis function expansion is that once the basis functions are determined, the regression analyses will be conducted in the vector space spanned by the unit vectors of transformed variables. Without having to rely on pre-determined parametric equations, extracted structural information can be interpreted in an unbiased manner.

We demonstrate the feasibility of our proposed approach through a scattering investigation of aqueous solutions of Pluronic L64, a well-studied micellar system.[24-25] Its chemical formula is $(PEO)_{13}(PPO)_{30}(PEO)_{13}$, where PEO stands for polyethylene oxide and PPO is polypropylene oxide. The molecular weight is 2990 Da. At low temperatures, both PEO and PPO are hydrophilic, so that L64 chains readily dissolve in water as monomers. Upon exposure to increased temperatures, hydrogen-bond formation between water and polymer molecules is perturbed and PPO tends to become less hydrophilic than PEO. As a result, an unbalance of hydrophilicity between the end blocks and the middle block of the polymer molecule is created. The copolymer molecules therefore acquire surfactant properties in the aqueous environment and self-assemble into micellar structures above the critical micellar temperature-concentration line. During the experiments, the temperature stability was controlled to within ± 0.1 °C. Samples with a fixed L64 concentration of 6 $wt\%$ were prepared by mixing a predetermined amount of $D_2O$ and de-ionized $H_2O$ with molar ratios of $D_2O$ to $D_2O + H_2O$, defined as $\gamma$, at 1, 0.9, and 0.8. Samples were prepared by weighing the appropriate mass of surfactant and adding the necessary volume of water with different values of $\gamma$ to achieve the desired concentrations. The prepared samples were further filtered by Anotop membrane filters (0.02 $\mu m$), mixed for 48 hours, and allowed to settle to equilibrium for 24 hours prior to the scattering experiments. SANS measurements were performed at the Extended Q-Range Small-Angle Neutron Scattering Diffractometer (EQ-SANS) at the Spallation Neutron Source (SNS), Oak Ridge National Laboratory (ORNL) and D22 - Large dynamic range small-angle diffractometer at Institute Laue-Langevin (ILL). The probed $Q$-range was from 0.003 to 0.4 $Å^{-1}$. Complementary ultra-small angle neutron scattering (USANS) measurement was carried out at USANS SNS to examine the potential existence of large-scale structure. The probed $Q$-range was from 0.0001 to 0.003 $Å^{-1}$. The samples for both SANS and USANS were accommodated in Hellma cells with a path length of 2 mm. The measured signal was corrected for detector background, sensitivity, and empty cell scattering, and were normalized by standard procedure in order to obtain absolute intensity.

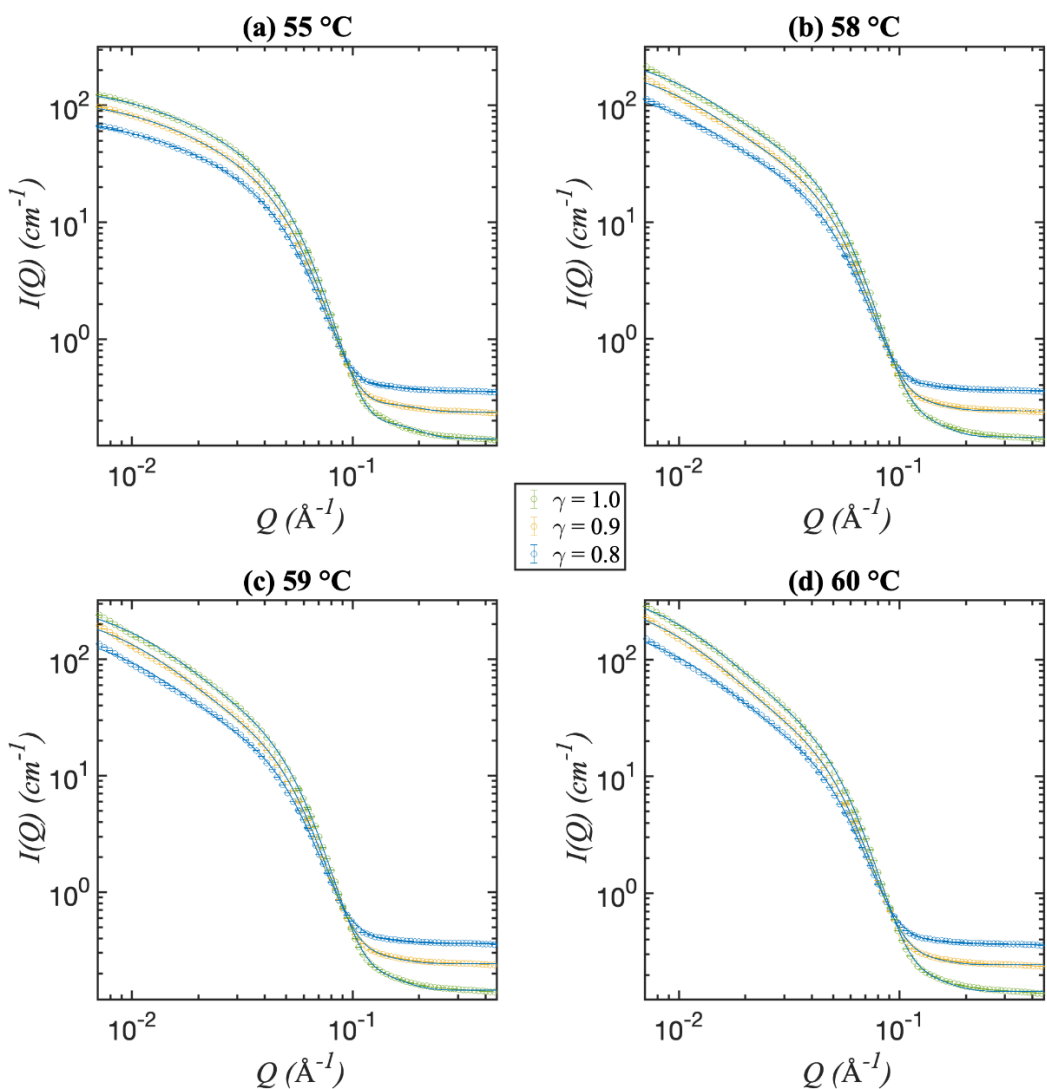

**Figure 1.** The SANS scattering cross section $I(Q)$ of L64 micellar solutions at 4 different temperatures for three different scattering contrasts. The experimental uncertainties are on the order or smaller than the symbol size. The solid lines are model fitting curves with the inclusion of instrumental resolution. Quantitative agreement between the experimental data and model curves is observed.

Before proceeding with further quantitative analysis, it is instructive to first examine the qualitative features of the measured SANS spectra. Fig. 1 gives the SANS contrast variation data obtained from the aqueous solutions of L64 with three with different values of $\gamma$. In agreement with the expectation, an increase in the molar ratio of H$_2$O in the solvent results in a monotonic decrease in the magnitude of the corresponding $I(Q)$ when $Q < 0.1$ Å$^{-1}$ and a steady increase in the incoherent scattering when $Q > 0.2$ Å$^{-1}$ due to the decreasing scattering contrast between L64 micelles and background water. In addition, within the probed temperature range of 55 °C to 60 °C, no perceptible correlation peak presents in the measured $I(Q)$. This observation suggests that the collected coherent scattering is essentially contributed by the intra-micellar spatial correlation. As a result, we did not consider the inter-micellar spatial correlation in our data analysis.

The intra-micellar scattering length density distributions of each constituent component of L64 micelles, including both (PEO)$_{13}$(PPO)$_{30}$(PEO)$_{13}$ building blocks and invasive water molecules, can be quantitatively evaluated using contrast variation SANS technique based on the model-free data analysis approach outlined below.

The measured SANS absolute intensity $I(Q)$ of L64 micellar solutions can be expressed by the following equation:[26]

$$I(Q) = n_m \Delta\rho_m^2 v_m^2 P_m(Q) + I_{inc} , (1)$$

where $I_{inc}$ is the incoherent background, which shows no dependence on $Q$, $n_m$ is the micellar number density, $\Delta\rho_m$ is the difference of the scattering length density (SLD) between micelles and background solvent, $v_m$ is the volume of a micelle, and $P_m(Q)$ is the form factor which describes the intra-micellar spatial correlation. $P_m(Q)$ is a normalized function with maxima of 1 at $Q = 0$. Similar to many amphiphiles in aqueous environments, the self-assembled structure of an L64 micelle is characterized by a densely packed core consisting of less hydrophilic PPO components surrounded by a more diffusive region composed of more polar PEO blocks. The density distribution of L64 molecules decays towards the periphery from the micellar core region. In addition, due to the difference in hydrophilicity of PEO and PPO, the statistical profile of water penetration is heterogeneous along the micellar radial direction: More invasive water molecules reside in the outer layer. Since the packing pattern of invasive water, in terms of two-point static correlation function, has been demonstrated to be different from that of bulk water, both polymer segment and invasive water molecules should be considered as the constituents of a L64 micelle. The quantity of interest is therefore $\Delta\rho_m$ since it is the only parameter in in Eqn. (1) which shows dependence on $\gamma$. One can be express $\Delta\rho_m$ as

$$\Delta\rho_m = \int_0^\infty \Delta\rho_m(r)\, 4\pi r^2 dr, (2)$$

where $\Delta\rho_m(r)$ is the radially averaged intra-micellar SLD distribution. Explicitly it can be expressed as

$$\Delta\rho_m(r) = \rho_p(r) + \rho_w(r) - \rho_w = \rho_p(r) + b_w\left[H(r) - \frac{1}{v_w}\right], (3)$$

where $\rho_p(r)$ and $\rho_w(r)$ are the radial SLD distributions of L64 molecules and invasive water, respectively. $H(r)$ is the radial number density distribution of invasive water. $\rho_w = \frac{b_w}{v_w}$ is the SLD of bulk water where $b_w$ is the scattering length of bulk water and $v_w$ is the volume of a water molecule, which is 30 Å$^3$. Assuming the isotopic effect is negligible and $\gamma$ of invasive water is identical to that of the bulk water, an contrast variation schme can be contemplated based on Eqn. (3) to determine $H(r)$ from the characteristic variation of $I(Q)$ given in Fig. 1: From the convolution theorem one can show that $n_m \Delta \rho_m^2 v_m^2 P_m(Q)$ in Eqn. (1) is the Fourier transform of the autocorrelation of $\Delta \rho_m(r)$. Namely,

$$\Delta \rho_m^2 v_m^2 P_m(Q) = \int \int dr dr' \Delta \rho_m(r) \Delta \rho_m(r') \exp\exp[-iQ \cdot (r-r')]. \quad (4)$$

By substituting $b_w = \gamma b_{D_2O} + (1-\gamma) b_{H_2O}$ into Eqn. (3) where $b_{D_2O}$ and $b_{H_2O}$ are bound scattering lengths of D₂O and H₂O respectively, a quantity $\sqrt{n_m}\Delta\rho_m^\gamma(r)$ can be defined by the following expression:

$$\sqrt{n_m}\Delta\rho_m^\gamma(r) = \gamma(b_{D_2O} - b_{H_2O})\sqrt{n_m}\left[H(r) - \frac{1}{v_w}\right] + \sqrt{n_m}\left\{\rho_p(r) + b_{H_2O}\left[H(r) - \frac{1}{v_w}\right]\right\} \quad (5)$$

From Eqn. (1) it is clear that $\sqrt{n_m}\Delta\rho_m^\gamma(r)$ can be obtained from the coherent component of measured $I(Q)$ as a function of $\gamma$. By differentiating Eqn. (5) with respect to $\gamma$, it is found that

$$\frac{d}{d\gamma}\left[\sqrt{n_m}\Delta\rho_m^\gamma(r)\right] = (b_{D_2O} - b_{H_2O})\sqrt{n_m}\left[H(r) - \frac{1}{v_w}\right] \quad (6)$$

Due to the nonpolar nature of PPO in aqueous environment, it is reasonable to assume that $H(0) = 0$. By imposing Eqn. (6) with sufficient boundary conditions, which can be achieved by altering $\gamma$, $H(r)$ can be uniquely determined. Moreover, since L64 surfactant molecules occupy all the intra-micellar space which is not accessible by invasive water, the radial number density distributions of L64 $P(r)$ can be determined by the following expression

$$P(r) = -\frac{\left[H(r) - \frac{1}{v_w}\right]v_w}{v_u}, \quad (7)$$

where $v_u$ is the molecular volume of a L64 surfactant. Based on its molecular weight ($2.9\ kg/mol$) and density ($1.07\ g/cm^3$), $v_u$ is found to be $4.5 \times 10^3$ Å³.

To select a proper spanning set used in the basis expansion for data analysis, one criterion is to examine the compatibility of boundary conditions inherent to the self-assembled structures and basis function. It is noticed that the asymptotic behaviors of Lorentzian function are indeed in agreement with two limiting properties of $H(r)$, namely $\lim_{r \to 0} \frac{dH(r)}{dr} = 0$ and $\lim_{r \to \infty} \frac{dH(r)}{dr} = 0$. Therefore, a complete set of Lorentzian-like function with $n$ member functions $u_n(r)$ is generated by the Gram–Schmidt process and used as the orthornormal basis vectors. Details of developing $u_n(x)$ are givin in the Supporting Information. Evidenced by Eqns. (4) and (5), $\Delta\rho_m^\gamma(r)$ is the quantity of interest in spectral analysis. As a starting point, $\Delta\rho_m^\gamma(r)$ is first expanded by $u_n(r)$. Namely,

$$\Delta\rho_m^\gamma(r) = \sum_{n=0}^{\infty} a_n^\gamma L_n(r), \quad (8)$$

where $a_n^\gamma$ are the associated expansion coefficients. From Eqns (1) one can further define $\Delta\rho_m^2 v_m^2 P_m(Q) \equiv [F_m(Q)]^2$. Therefore $F_m^\gamma(Q)$ is found to be

$$F_m^\gamma(Q) \equiv 4\pi \int_0^\infty dr r^2 \Delta\rho_m^\gamma(r) j_0(Qr) = \sum_{n=1}^{\infty} b_n^\gamma v_n(Q) \quad (9)$$

where $v_n(Q) = \sqrt{\frac{2}{\pi}} \int_0^\infty dr r^2 u_n(r) j_0(Qr)$, $u_n(r) = \sqrt{\frac{2}{\pi}} \int_0^\infty dQ Q^2 v_n(Q) j_0(Qr)$, and $j_0(x)$ is the Bessel function of the first kind of order zero. $b_n^\gamma$ are linearly proportonal to $a_n^\gamma$. From Eqns. (1), (4), and (8), $b_n^\gamma$ can be determined from measured spectra and accordingly $\Delta\rho_m^\gamma(r)$ can be inversely reconstructed via Eqn. (8). An example of implementation of the regression analysis based on our proposed basis expansion strategy is given in Fig. 2: The experimentally measured $I(Q)$ given in Fig. 2(a) can be well approximated by a series of 5 basis functions. The disagreement is found to be within the statistical errors. From the associated expansion coefficients given in Fig. 2(b), the quantity $\sqrt{n_m}\Delta\rho_m^\gamma(r)$ can be reliably extxtracted without using a predetermined expression as input. The merit of our proposed approach can be demonstrated by comparing with existing numerical procedures developed for extracting structural information from experimental small angle scattering data, such as indirect Fourier transformation:[27-28] First, the basis functions selected in our inverse scheme form a complete set, which ensures stable solutions in regression analysis. Since our selected basis functions are also compatible with boundary conditions inherent to the structure of studied systems, faster convergency in determining the coefficients associated with the related basis functions is guaranteed. Moreover, the numerical errors in the optimization process of spectral analysis can be significantly minimized since our selected basis functions can be expressed analytically.

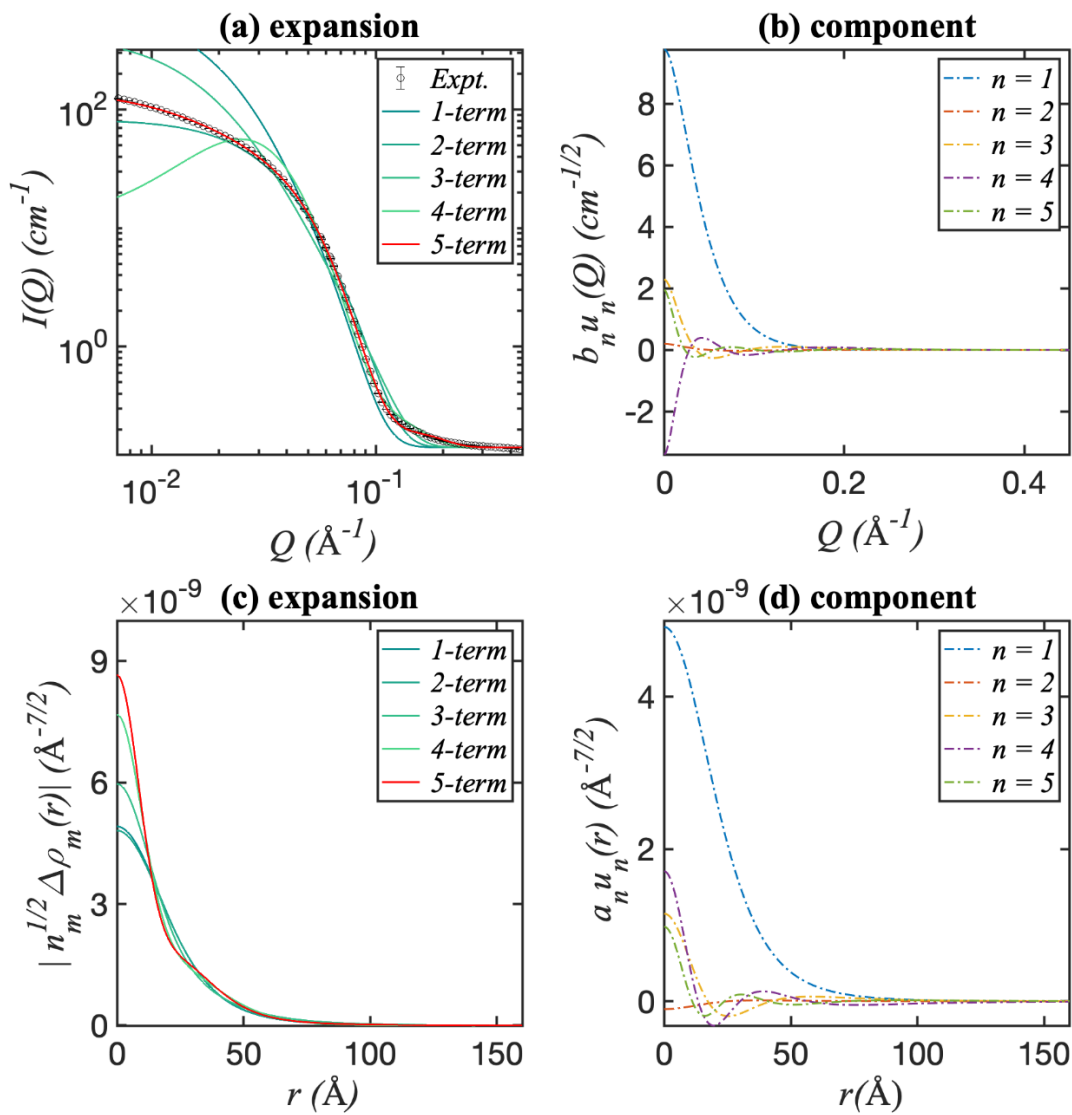

**Figure 2.** (a) The $I(Q)$ corresponding to the L64 micellar solution at 55 °C with $\gamma = 1$ (black symbols) can be well approximated by a series of 5 basis functions (red solid curve). The disagreement is within experimental error. (b) The expansion decomposition from the corresponding five basis functions of $b_n^\gamma v_n(Q)$. (c) The extracted $\sqrt{n_m}\Delta\rho_m(r)$ is given by the red solid curve. (d) The expansion decomposition from the corresponding five basis functions of $a_n^\gamma u_n(r)$. It is clear that $a_n^\gamma$ are linearly proportional to $b_n^\gamma$.

In Fig. 3 we present the extracted $\sqrt{n_m}\Delta\rho_m(r)$ corresponding to the various $I(Q)$ given in Fig. 1. The results are further multiplied by $r^2$ to demonstrate detailed features. Qualitatively, the radial distribution of $\Delta\rho_m(r)$ can be divided into three regions: The central core region ranging from $r = 0$ to approximately $r = 20$ Å, the intermediate region where a characterostic peak is observed between $r = 20$ Å and $r = 80$ Å, and the micellar periphery region when $r > 80$ Å. As indicated by Eqn. (3), it can be inferred that the observed fluctuation of SLD distribution reflects the interplay of $H(r)$ and $P(r)$. Moreover, because of the nonparametric nature of our proposed approach, the radially averaged characteristic variation of $\sqrt{n_m}\Delta\rho_m(r)$ displayed in Fig. 3 is not artifically generated by the imposed deterministic function as in existing paranetric approaches.

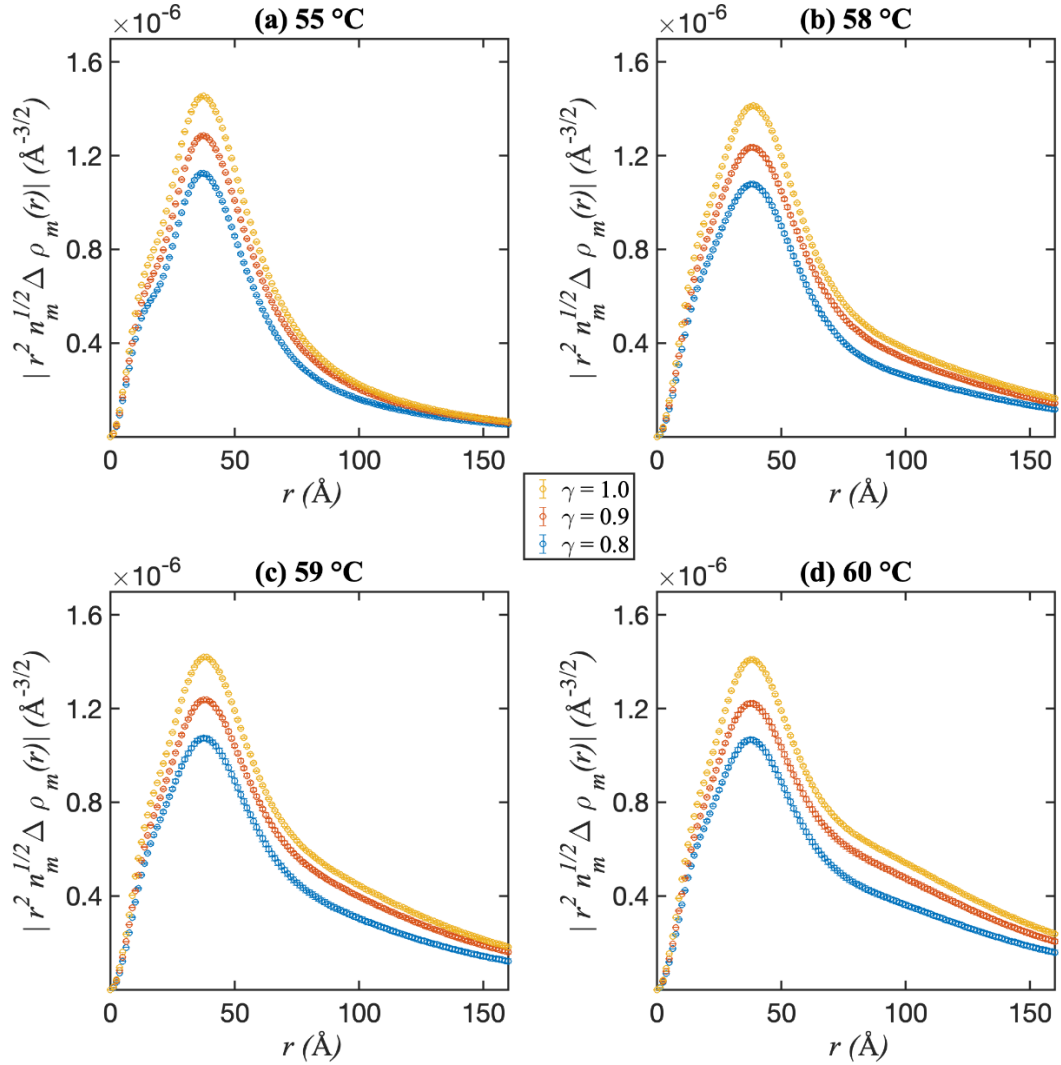

**Figure 3.** $|r^2\sqrt{n_m}\Delta\rho_m(r)|$ as a function of $\gamma$ corresponding to the $I(Q)$ given in Figure 1. With the probed temperature range, $|r^2\sqrt{n_m}\Delta\rho_m(r)|$ can be qualitatively separated into three different regions.

From the extracted $r^2\sqrt{n_m}\Delta\rho_m(r)$ at different $\gamma$ given in Fig. 3, we can readily determine $H(r)$ and $P(r)$ from Eqns. (6) & (7) and the results are given in Panels (a) and (b) of Fig. 4. Within the probed temperature range of 55 °C to 60 °C, $H(r)$ is seen to approach its mean-field limit of $\frac{1}{v_w}$ $\left(\frac{1}{30}\ \text{Å}^{-3}\right)$ when $r \gtrsim 100$ Å. As indicated by the inset of Fig. 4(a), a steady and discernible decrease of $H(r)$ is observed in the spatial region of 80 Å $< r <$ 120 Å. $\Delta H(r)$ is defined as the difference between $H(r)$ and $H(r)$ at 55 °C. This observed dehydration in the micellar peripheral region is also revealed by the extracted $P(r)$ given in Fig. 4(b): Indicated by the variation of $P(r)$ in the region of $r > 80$ Å, it is clear that the L64 micelle grows as temperature increases from 55 °C to 60 °C because of the increased hydrophobicity of the polymer segments at higher temperatures.

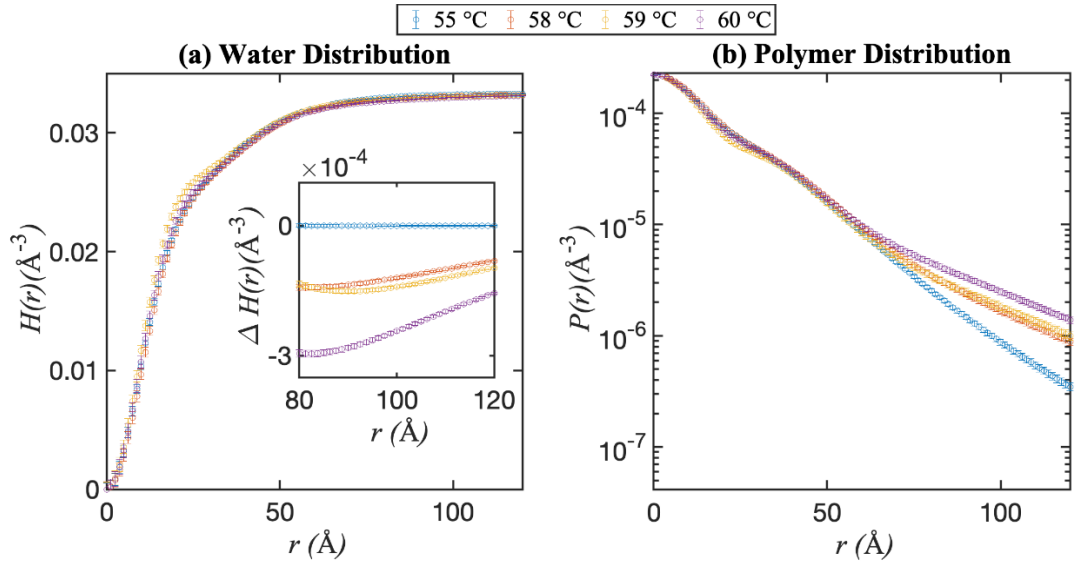

**Figure 4.** (a) The extracted radial distributions of the intra-micellar water and (b) number density L64 at different temperatures.

The observed enhanced self-association can be quantitatively measured by the averaged micellar aggregation number and radius. From $H(r)$, the intra-micellar space $v_{ew}$, which is not accessible by water, can be defined as

$$v_{ew} \equiv -v_w \int_0^\infty \left[ H(r) - \frac{1}{v_w} \right] 4\pi r^2 dr, (10)$$

Accordingly, $N_{agg}$ can be defined as

$$N_{agg} = \frac{v_{ew}}{v_u}, (11)$$

Obtained calculations of $N_{agg}$ as determined from Eqns. (10) and (11) are displayed in Fig. 5(a). $N_{agg}$ increases from 35 to 60 upon increasing the solution temperature from 55 °C to 60 °C. The micellar size can be calculated from $N_{agg}$ and $P(r)$: By imposing the constraint that L64 surfactant molecules only exist as unimers in water, a spatial threshold $R_b$ can be defined by the following equation:

$$\int_0^{R_b} P(r) 4\pi r^2 dr = N_{agg} - 1, (12)$$

Obtained values of $R_b$ calculated from Eqn. (12) are given in Fig. 5(b). The steady increase in $R_b$ again reflects the increasing unbalance of hydrophilicity between the PEO end blocks and the PPO middle block of L64 molecules at higher temperatures. Moreover, we further calculate the radius of gyration $R_g$ through the average of $r^2$ with respect to $P(r)$ and give the results in Fig. 5(c). It is instructive to discuss the comparison between $R_b$ and $R_g$: $R_b$ is larger than $R_g$ by a factor of 2 to 3. As a quantitative measure of micellar size, $R_b$ is more sensitive to the conformational change since $R_g$ only detects the change in the second moment of $P(r)$. An interesting view of the micellar conformation may be obtained by inspecting the ratio of $\frac{R_b}{R_g}$ as a function of temperature. For a hard sphere particle with uniform density profile, it is 1.29 since $R_g = \sqrt{\frac{3}{5}} R_b$. For an ideal hollow particle whose mass distribution is localized at the infinitely thin shell, the value of $\frac{R_b}{R_g}$ can be calculated to be 1. Within the probed temperature range, the values of $\frac{R_b}{R_g}$ are found to be substantially larger than that for the monodispersed hard sphere, 1.29. This observation suggests that the traditional "dense-core" picture of L64 micelles is flawed. Moreover, the steady increase of $\frac{R_b}{R_g}$ again manifests the increase in hydrophobic character of L64 molecules in aqueous solutions upon raising temperature.

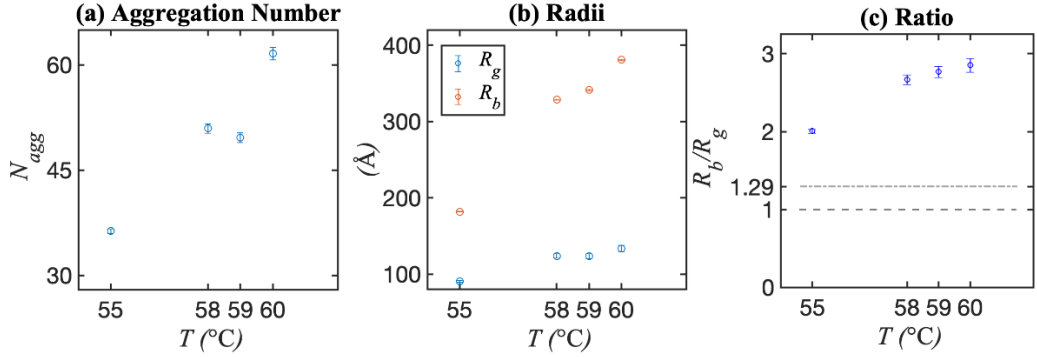

**Figure 5.** (a) The averaged micellar aggregation number $N_{agg}$ (b) Radius of gyration $R_g$ (blue circles) and micellar boundary $R_b$ (orange circles) (c) The ratio of $R_b$ to $R_g$ as a function of temperature. The dotted line and dashed line give the values of $\frac{R_b}{R_g}$ for hard sphere and hollow sphere respectively.

In summary, we have developed a model-free approach, based on the scheme of basis function expansion, to quantitatively evaluate the structure of micellar solutions from the coherent component of their small angle neutron scattering intensities. Complemented by the scheme of contrast variation, we show the conformational characteristics of micelles can be determined without having to rely on pre-determined model which could cause potential issue of biased interpretation of experimental data. Comparisons to existing numerical procedures demonstrate the merits of our proposed method in accuracy, efficiency, and general applicability.

The prospect of our proposed approach for scattering data analysis appears to be appealing: It can be applied to examine the influence of invasive solvent molecules on conformational flexibility in a variety of globular colloids with molecular architecture open for solvent penetration. Self-assembled systems whose structures lack spherical symmetry, such as wormlike and lamellar micelles, can also be addressed based on the extension of same expansion strategy using basis functions with suitable spatial symmetry.

## ACKNOWLEDGMENTS

This research was performed at SNS, which is a DOE Office of Science User Facility operated by the Oak Ridge National Laboratory. A. H. acknowledges the support of the ORNL HSRE program. We greatly appreciate the EQ-SANS beamtime at the SNS.